\documentclass[10pt,letterpaper]{article}

\usepackage[final]{cogsci}
\usepackage{graphicx}
\usepackage{caption}

\usepackage{url}

\usepackage[
  style=apa,
  natbib=true,
  annotation=false,
]{biblatex}
\usepackage{float} %%% Roger Levy added this and changed figure/table placement to [H] for conformity to Word template, though floating tables and figures to top is still generally recommended!

\title{An Experimental Method to Study Opinion Diffusion in Human-AI Hybrid Societies} 

\author[1]{\mbox{Léna Gaubert}} % lena@arlq.ai
\author[1]{\mbox{Rémi Devaux}} % remi@arlq.ai
\author[2]{\mbox{Elif Çelen}} % elif.celen@ae.mpg.de
\author[3]{\mbox{Raja Marjieh}} % raja.marjieh@princeton.edu
\author[4]{\mbox{Diana Mangalagiu}} % diana.mangalagiu@sbs.ox.ac.uk
\author[1]{\mbox{Antoine Jardin}} % antoine@arlq.ai
\author[5]{\mbox{Nori Jacoby}} % kj338@cornell.edu

\affil[1]{Arlequin AI, Paris, France}
\affil[2]{Max Planck Institute for Empirical Aesthetics, Frankfurt, Germany}
\affil[3]{Department of Psychology, Princeton University, Princeton, New Jersey, United States}
\affil[4]{Saïd Business School, University of Oxford, Oxford, United Kingdom}
\affil[5]{Department of Psychology, Cornell University, Ithaca, New York, United States}

\begin{document}

\maketitle

\begin{abstract}

As artificial intelligence increasingly mediates public discourse, it becomes important to understand how human-AI collectives shape opinion formation, deliberation, and democratic outcomes. We present a novel experimental method for studying opinion dynamics in hybrid human-AI social networks. Participants, human or AI, were embedded in 5×5 grid lattice networks and iteratively asked to select and revise statements on a given polarizing topic over eight rounds. We compared three conditions: human-only, AI-only, and hybrid networks with equal proportions of human and AI participants. Hybrid human-AI networks achieved the lowest final polarization while, in contrast, human-only networks exhibited higher polarization with lower neighbor agreement. We also ran additional experiments varying Large Language Model (LLM) prompt framing to explore whether instruction design might influence convergence patterns. Although these early findings are preliminary and cannot yet support broad generalizations, they highlight the potential value of experimental social networks for understanding opinion dynamics in human-AI hybrid societies.

\end{abstract}
%%% FIGURE 1 – SCHEMATICS
\begin{figure*}[h!]
    \begin{center}
    \includegraphics[width=\textwidth]{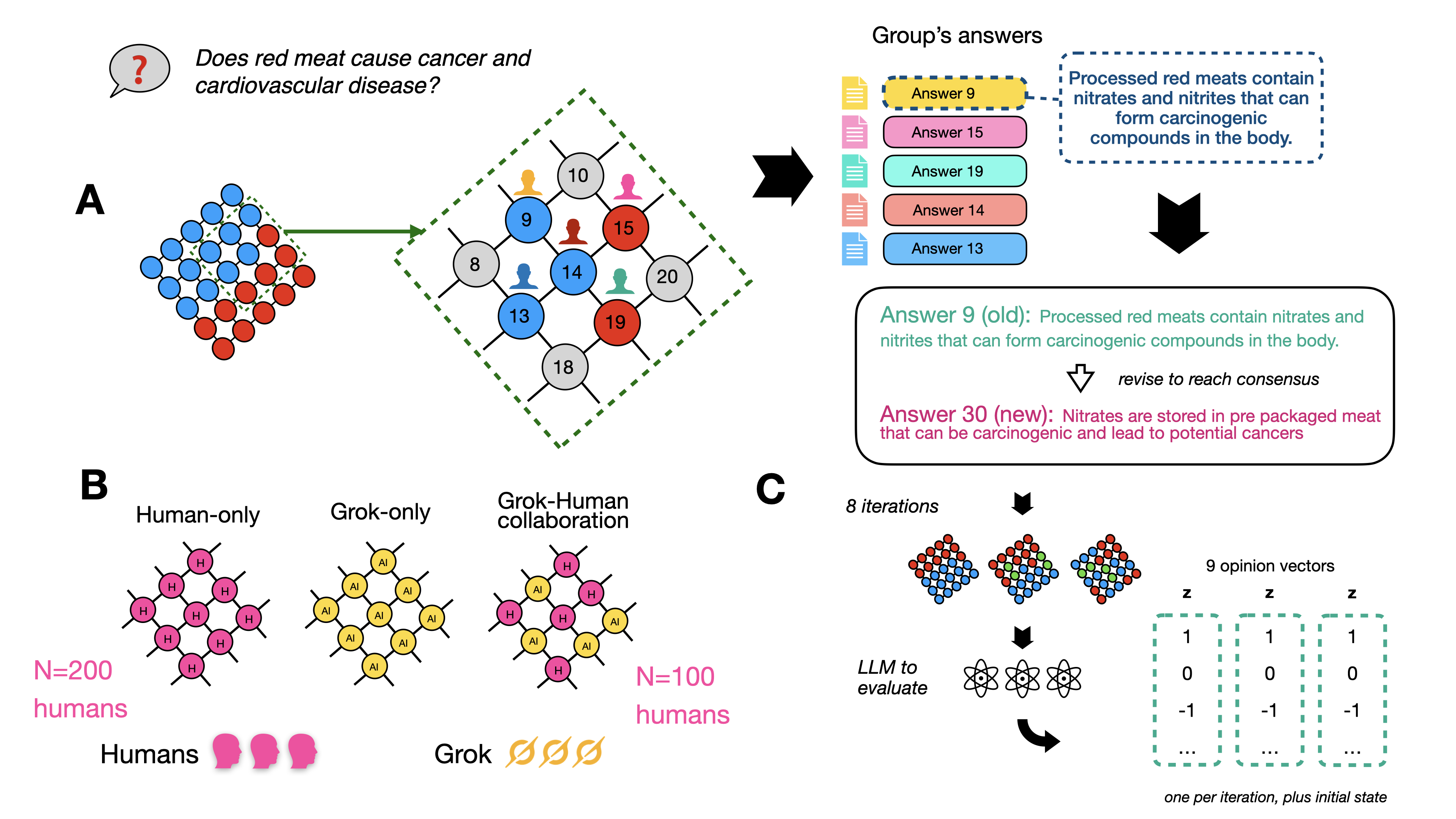}
    \caption{\textbf{Experimental design.}
    \textbf{A.} Humans or AI agents were embedded in a $5 \times 5$ social network that evolved over eight iterations. The network was initialized with positive opinions in the upper half and negative opinions in the lower half (supporting vs.\ not supporting the central claim). In each trial, participants viewed three to five opinions from neighboring nodes in the previous iteration (up, down, left, right, and the prior node at the same location) and selected the statement they felt best represented the group. They were then asked to rephrase the statement to be consistent with prior observations.
    \textbf{B.} We tested three conditions: human-only, AI-only (Grok), and hybrid (human–AI). \textbf{C.} An LLM was used to annotate each statement for positivity, negativity, and neutrality.}
    \label{fig-1:schematics}
    \end{center}
\end{figure*}

\section{Introduction}

The exchange of opinions has consistently shaped how societies form consensus, navigate disagreement, and make collective decisions \citep{acemoglu2011opinion}. Artificial intelligence (AI) increasingly mediates these exchanges as active participants generate content, curate information, and engage directly in public discourse \citep{brinkmann2023machine}. In this emerging landscape, networks in which AI-agents and humans interact constitute complex social systems whose collective dynamics cannot be predicted from the behavior of either alone \citep{tsvetkova2024new}. Understanding how opinion formation unfolds in these hybrid networks has become a challenge with profound implications for opinion dynamics, democratic outcomes and social cohesion \citep{tessler2024ai}.

Algorithmic recommendation systems have been shown to create filter bubbles that reinforce existing beliefs \citep{chitra_analyzing_2020} while the deployment of AI-generated content on platforms such as X raises new questions about how synthetic voices shape public opinion at scale. However, the picture is not uniformly bleak: recent evidence suggests that AI can outperform human mediators in facilitating convergence among individuals with divergent views, helping groups find common ground while preserving minority perspectives \citep{tessler_ai_2024}. More broadly, AI can enhance group performance in a variety of tasks such as cognitive modeling \citep{binz2025foundation,rmus2025generating}, creativity \citep{doshi2024generative,breithaupt2024humans}, and  scientific discovery \citep{jumper2021highly,schmidgall2025agent}.  These contrasting findings highlight a fundamental gap in understanding how the composition of human-AI collectives influences the trajectory of opinion dynamics, and whether hybrid networks amplify or attenuate polarization.

Our novel approach seeks to address this gap through controlled experiments that embed human and AI participants within social networks, enabling systematic comparison across network compositions. Our preliminary findings hint at possible links between network composition and polarization and opinion trajectories. 

\section{Background}

\subsection{AI and Collective Dynamics}

As AI systems become embedded in human activity, their impact extends beyond individuals to shape collective behavior and social dynamics at scale \citep{tsvetkova2024new}. Large Language Models (LLMs) not only impact individual cognition, but also collective cognition, generating new interactions structures between multiple individuals and AI agents simultaneously \citep{sucholutsky2025using}. Researchers have begun deploying LLMs in multiple roles within collective settings to study how human-AI interaction transforms group-level outcomes \citep{tessler_ai_2024, donkers_human-agent_2025, shiiku_dynamics_2025,hashemi2025collective, williams2023epidemic}.

Recent experimental work reveals that hybrid human-AI collectives exhibit dynamics distinct from purely human or purely AI groups \citep{shiiku_dynamics_2025}. Network structure further moderates these dynamics, with graph topology shaping both the semantics of shared narratives and the personal representations of individuals in experimental social networks \citep{priniski_network_2026}.

\subsection{AI Collectives and Democratic Outcomes}

The widespread adoption of AI agents, as content generation tools and social media users (e.g. X’s Grok) has direct implications for democratic processes. Experimental evidence investigates the effect of \textit{opinion cascades} on polarization between partisans \citep{macy_opinion_2019}. Once initiated, polarization tends to intensify over time, as new topics become increasingly biased \citep{zhao_evolution_2023, pournaki_conflicting_2025}. Algorithmic filtering further amplifies these dynamics by connecting users with content they already agree with (\textit{echo-chambers}, \cite{chitra_analyzing_2020}).

LLM-based frameworks now enable researchers to study these phenomena with both analytical precision and linguistic realism, reproducing echo-chamber dynamics while testing mitigation strategies in controlled settings \citep{wang_decoding_2024, donkers_human-agent_2025}. Beyond polarization, AI can outperform human mediators in bridging divergent opinions, facilitating convergence while incorporating dissenting voices \citep{tessler_ai_2024}.

\subsection{Polarization Models}

As AI systems increasingly mediate online discourse through recommendation algorithms and content generation, measuring and modeling polarization in human-AI environments has become a central challenge. Quantifying polarization requires methods that capture both the structural and dynamic properties of opinion formation in social networks \citep{garimella_quantifying_2018}. Metrics such as Polarization Index \citep{matakos_measuring_2017}, Neighbors Correlation Index (NCI) \citep{cinus_effect_2022} score are widely used across polarization studies.

Classical frameworks such as the Friedkin-Johnsen model and the Bounded Confidence Model formalize opinion updating through social influence \citep{friedkin1990social, deffuant2000mixing}. Extensions to these models have incorporated algorithmic filtering to capture platform-level effects on polarization \citep{chitra_analyzing_2020}, while other formal models that integrate homophily and social balance reveal phase transitions where increased connectivity triggers explosive polarization \citep{thurner_why_2025}.

However, these approaches represent opinions numerically, neglecting the textual and semantic nature of real communication. Recent work addresses this limitation through LLM-based simulations that embed formal opinion dynamics principles within language-capable agents \citep{wang_decoding_2024}.
%%% FIGURE 2 – ANOTATIONS
\begin{figure*}[h!]
    \begin{center}
        \includegraphics[width=0.9\textwidth]{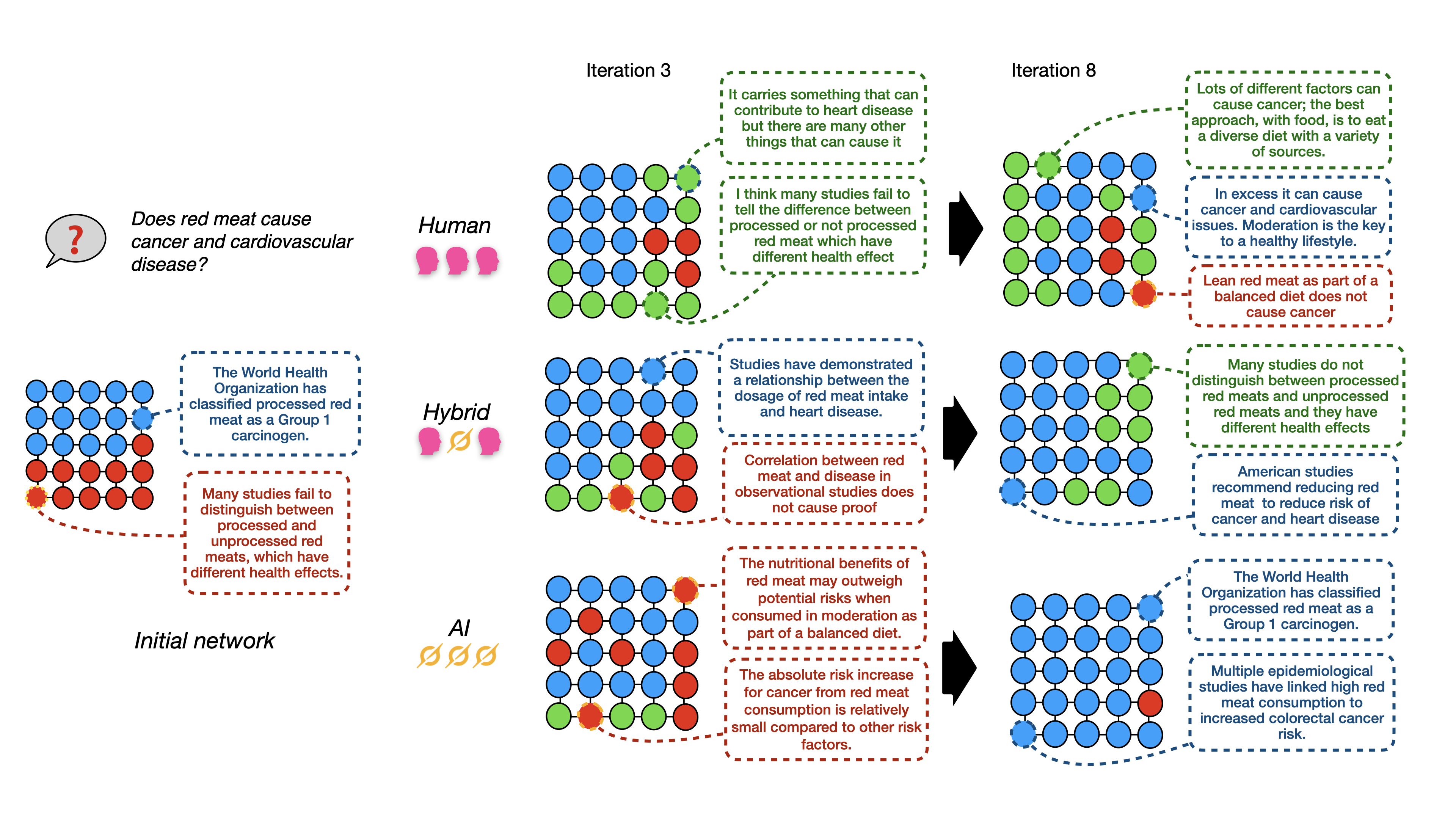}
        \caption{\textbf{Example statements.} Example statements from the three conditions at the third and eighth iterations, shown.}
        \label{fig-2:AI-annotations}
    \end{center}
\end{figure*}
\section{Method}

\subsection{Experiment Design}
\label{part: experimental_design}

We conducted large-scale online experiments in which participants (human, hybrid, or fully AI) were embedded in a directed unweighted 5×5 grid lattice social network (25 nodes; figure \ref{fig-1:schematics}A). All networks were initialized with the question: \textit{Does red meat cause cancer and cardiovascular disease?} The question was chosen to be engaging while intentionally avoiding highly divisive political content, in order to minimize polarization driven by pre-existing opinions. For this question, we generated a set of 24 statements: 12 positive and 12 negative. The positive statements supported an affirmative response to the question, while the negative statements argued against the proposed claim. Statements were generated using Claude 3.7 Sonnet and stored in a JSON database used to initialize the experiments' networks.
In each network, the 25 nodes were populated with samples drawn from this statement set. At the initial stage (iteration 0), nodes were assigned statements such that no two neighboring nodes shared the same statement, while maintaining a slight imbalance in favor of one opinion (e.g., 14 positive and 11 negative; see Figure \ref{fig-1:schematics}A). At each iteration, the experiment proceeded as follows:

\begin{enumerate}
    \item Participants were presented with the question and a list of statements from their  neighboring nodes in the previous iteration (up, down, left, right, and the prior node at the same location) in the network and asked to ``\textit{choose the answer this group would most likely agree with}''.
    \item After a one-minute display of their choice, participants were asked to revise their answer ``\textit{so that it most accurately reflects the views that the previously observed group of participants would be likely to agree with.}''.
    \item Participants then entered their revised response. Participants were prevented from switching tabs or opening new windows during the experiment, and we checked that each response contained at least 5 words.
\end{enumerate}

AI participants received similar instructions, with additional context provided (the experiment setup was made explicit, and the LLM was shown the previously observed group's opinions when revising answers) and guidance to ensure task completion (no justifications were required when choosing answers).

After a participant interacted with the opinions associated with a given node and submitted a revised statement, that statement became the updated state of the node. The participant was then dismissed, and a new participant was recruited and assigned to another node. In this way, statements were transmitted and revised throughout the network. Overall, we completed eight full updates of all 25 nodes, corresponding to eight iterations. To minimize bias, participants were not informed whether their neighbors’ statements were generated by humans or by AI. 
Experiments were implemented with PsyNet \citep{harrison2020gibbs}, a Python-based framework for advanced behavioral experiments and social simulations, that ensures that participants assigned to a given network position were recruited only after all neighboring positions from previous iterations had been filled. The experiments were inspired by prior experimental social network designs proposed by \citet{shiiku_dynamics_2025,marjieh2025characterizing}.

\begin{figure*}[h!]
    \begin{center}
    \includegraphics[width=0.85\textwidth]{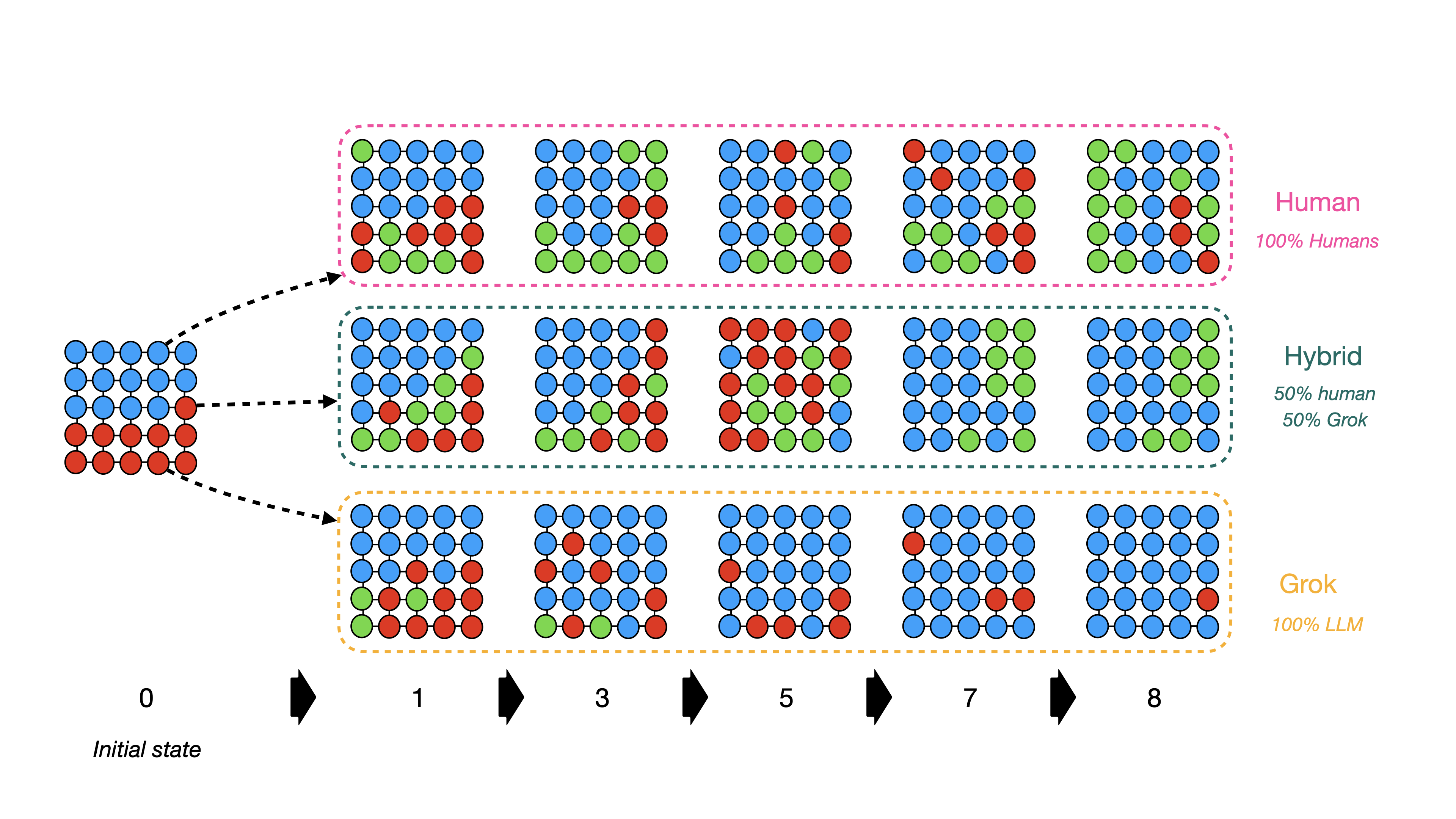}
    \captionsetup{skip=2pt}
   \caption{\textbf{Example trajectories.} Statements are annotated by an LLM as positive (blue), negative (red), or neutral (green) in regards to the given question.}
   
    \label{fig-3:grids-examples}
    \end{center}
\end{figure*}

\subsection{Experimental Conditions, Participants, and AI  calls}

We compared three experimental conditions (Figure~\ref{fig-1:schematics}B): 
(1) \textbf{Human-only}, in which all network nodes were occupied by human participants ($N=200$); 
(2) \textbf{AI-only}, in which all nodes were populated by Grok-4-Fast agents (42 runs; 400 API calls per run; total 16,800); and 
(3) \textbf{Hybrid}, consisting of an equal mix of human participants ($N=100$) and Grok-4-Fast agents (50\%; 200 model calls). 

\textbf{Human participants} were recruited online via Prolific\footnote{\url{https://www.prolific.com}}. All participants were based in the UK and reported English as their native language.  Participants provided informed under a Cornell University-approved protocol (IRB0148995) and were compensated at a rate of \$9 per hour. In total, 300 participants were recruited.  %excluding failed nodes

We chose Grok as our AI participant due to its training on real online social interactions and its active integration within a social media platform, making it an interesting choice for modeling AI behavior in networked social environments (see Limits \& Future Directions). \textbf{AI-generated} statements were thus produced using Grok-4-Fast via OpenRouter's API\footnote{\url{https://openrouter.ai/}}. Totaling, 17{,}000 API calls.  % 16,800 full-AI + 200 hybrid

All collected data, statement database and prompts are available on Gitlab at the following repository: \url{https://gitlab.com/nori.jacoby/hybrid-opinion-data}.

\subsection{Measures of Polarization}
Figure \ref{fig-1:schematics}C, shows the schematics of the annotation process. For each node at each iteration, we collected the participant's revised statement. We follow the opinion dynamics literature by denoting node $i$’s opinion as $z_i \in \{-1,0,1\}$. Each node's opinion variable is coded based on their revised answer to the question: 1 for arguments supporting a positive response, $-$1 for arguments supporting a negative response, and 0 for neutral answers or those presenting both perspectives.

Opinion variables were inferred by presenting each node's answers in batches of 20 to Claude Sonnet 3.7, which annotated them as negative, neutral, or positive with respect to the original question.

We then used the participants’ resulting opinions to compute two popular polarization metrics.

\paragraph{Polarization Index \citep{matakos_measuring_2017}} For directed unweighted networks—where participants' \textit{innate} opinions are unknown, polarization $P_z$ can be defined as the variance of \textit{expressed} opinions across the network at a given iteration:

$$
P_z = \frac{1}{N} \sum_{i=1}^{N} (z_i-\bar{z})^2,
$$

where $\bar{z}$ represents the average \textit{expressed} opinion in the network. This metric quantifies the extent to which opinions vary within the network. 

\paragraph{Neighbors Correlation Index (NCI, \cite{cinus_effect_2022})} The NCI is calculated using the opinion vector $\mathbf{z} = [z_1, z_2,...z_N]$, which contains the opinion of each participant. It is defined as the correlation between the opinion vector of each participant and the neighbors' opinion vector $\mathbf{n} = [n_1, n_2,...n_N]$ which contains the average opinion of each participant's neighbors. Denoting participant $i$'s set of neighbors by $\mathcal{N}_i$, the $i$-\textit{th} element of $\mathbf{n}$ is defined as the average of $i$'s neighbors opinion:
$$
n_i = \frac{1}{|\mathcal{N}_i|} \sum_{j \in \mathcal{N}_i}z_j.
$$
The NCI is then defined as the correlation between the two vectors: $
\text{NCI} = \rho(\mathbf{z}, \mathbf{n})$, where $\rho(.)$ denotes the Pearson correlation function. Thus, the NCI is defined on the interval ($-$1,1) and measures the degree to which participants agree with their neighbors within the network. An NCI of 1 indicates that all participants in the network agree.

\section{Preliminary Results}

In this section, we present the results obtained across the three experimental conditions (human-only, AI-only, and human-AI hybrid). These results are preliminary, as additional experiments across a broader set of questions are required to generalize our findings and provide further statistical support for the trends we observed in the current dataset. 

\subsection{Opinion Dynamics across Human, AI and Hybrid Networks}

Figure \ref{fig-2:AI-annotations} shows examples of statements produced by humans and AI in the early iterations. Human statements initially contained personal opinions (e.g., “I think…”) and vague language (e.g., “it carries something…”), but became more refined in later iterations, converging toward formulations similar to common public discourse (e.g., “the best approach is to eat a diverse diet…”). AI-generated statements were more detailed from the beginning, frequently invoking authority (e.g., the World Health Organization) and technical terminology (e.g., “colorectal cancer”). Statements in the hybrid condition exhibited an intermediate style, combining elements of both approaches—for example, referencing studies in a more colloquial register. Together, these observations suggest that humans and AI initially employ different communicative strategies.

Figure~\ref{fig-3:grids-examples} shows example opinion trajectories across the three conditions. In all conditions, opinions form clusters with similar annotations. The human-only condition shows the greatest diversity of opinions in the final iteration, followed by the AI-only condition. In the hybrid condition, negative opinions disappear over time, with positive and neutral statements becoming dominant.

\begin{figure}[ht]
  \begin{center}
    \includegraphics[width=0.45\textwidth]{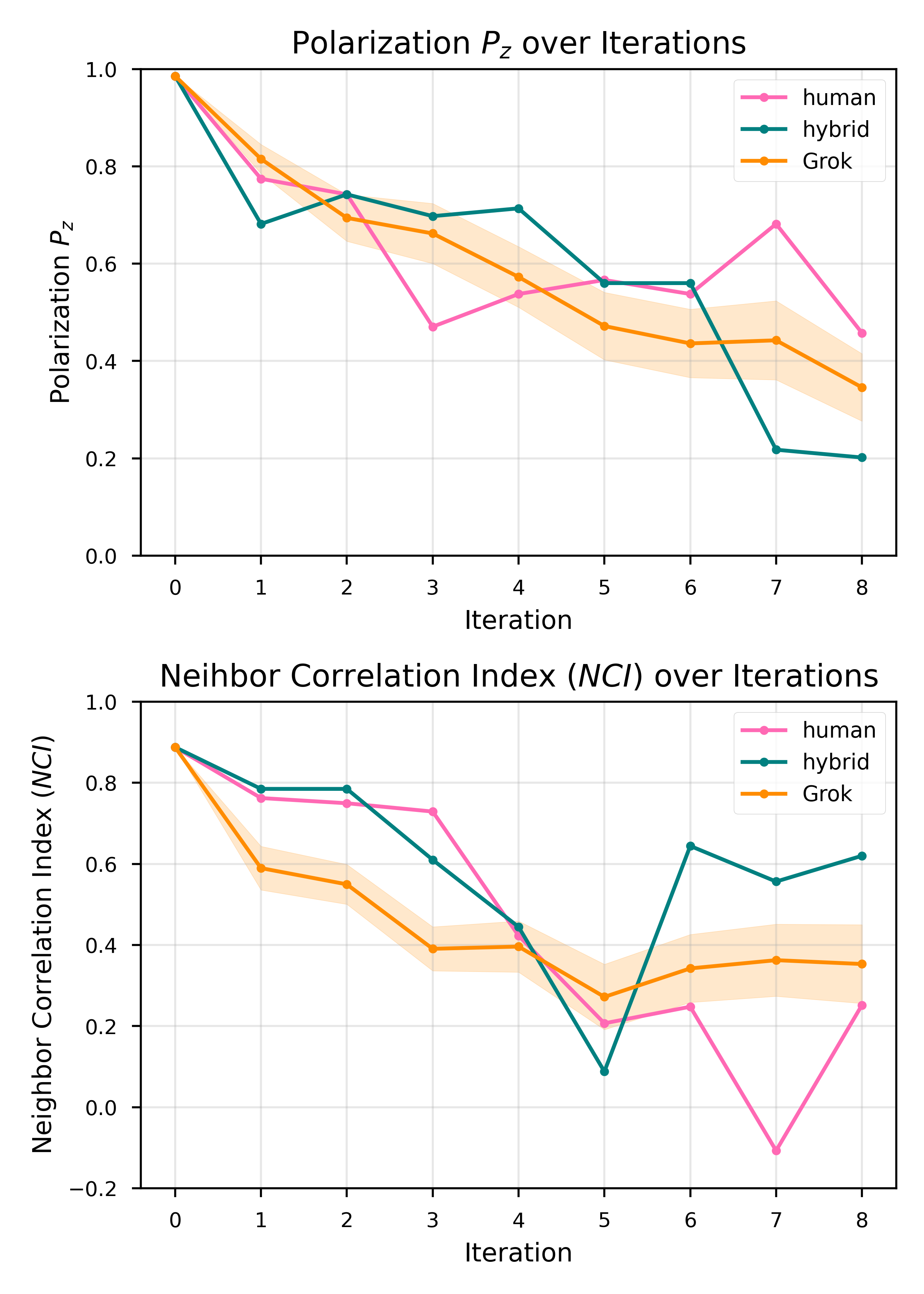}
 
  \captionsetup{skip=2pt}
  \caption{\textbf{Experimental results}. Evolution of Polarization (top) and Neighbors Correlation Index (bottom) over iterations across the three experimental conditions: \textbf{human-only} (pink), \textbf{Grok-only} (dark orange) and \textbf{hybrid} (steel blue). Shaded area (Grok-only) represents $+/-1$ standard error of the mean across 21 runs.}
  \label{fig-4:main_results}
   \end{center}
\end{figure}

Figure \ref{fig-4:main_results} presents the evolution of polarization ($P_z$) and NCI over iterations for the three experimental conditions. The AI only condition was run 21 times where the human and hybrid experiment were run only once. All conditions exhibited an initial decrease in polarization, indicating opinion convergence over the course of the experiment. Changes in NCI were more variable, but followed a similar overall downward trend. For the AI-only condition, where statistical tests was possible across repeated runs, both $P_z$ and NCI decreased significantly ($P_z$: $r = -0.97$,  $CI_{95} =[-0.99, -0.95]$; NCI: $r = -0.80$, $CI_{95} =[-0.95, -0.44]$).

The human-only condition yielded the highest final polarization score ($P_z=0.45$). The gradual decline in polarization, combined with a substantial decrease in NCI (final $\text{NCI}=0.13$), suggests gradual diffusion and mixing of opinions across the network, resulting in global disagreement among neighbors. In the LLM-only (Grok) condition, final values averaged $M=0.36$, ($CI_{95}=[0.22, 0.49]$) for polarization and $M=0.33$ ($CI_{95}=[0.13, 0.51]$) for NCI. In particular, the hybrid condition achieved the lowest final polarization ($0.20$) alongside the highest final NCI ($0.61$), suggesting that human-LLM interactions may accelerate depolarization, though through a less predictable process. These preliminary results suggest that the presence of LLMs may differentially influence opinion dynamics as a function of network composition. In particular, hybrid human–AI networks appear to converge more rapidly and to exhibit more clustered spatial organization, consistent with the interpretation that LLMs can act as mediators that accelerate consensus formation among human participants \citep{tessler_ai_2024}. However, additional data—specifically, repeated experimental runs—are required to assess the statistical robustness of these effects.

\begin{figure}[ht]
  \begin{center}
    \includegraphics[width=0.45\textwidth]{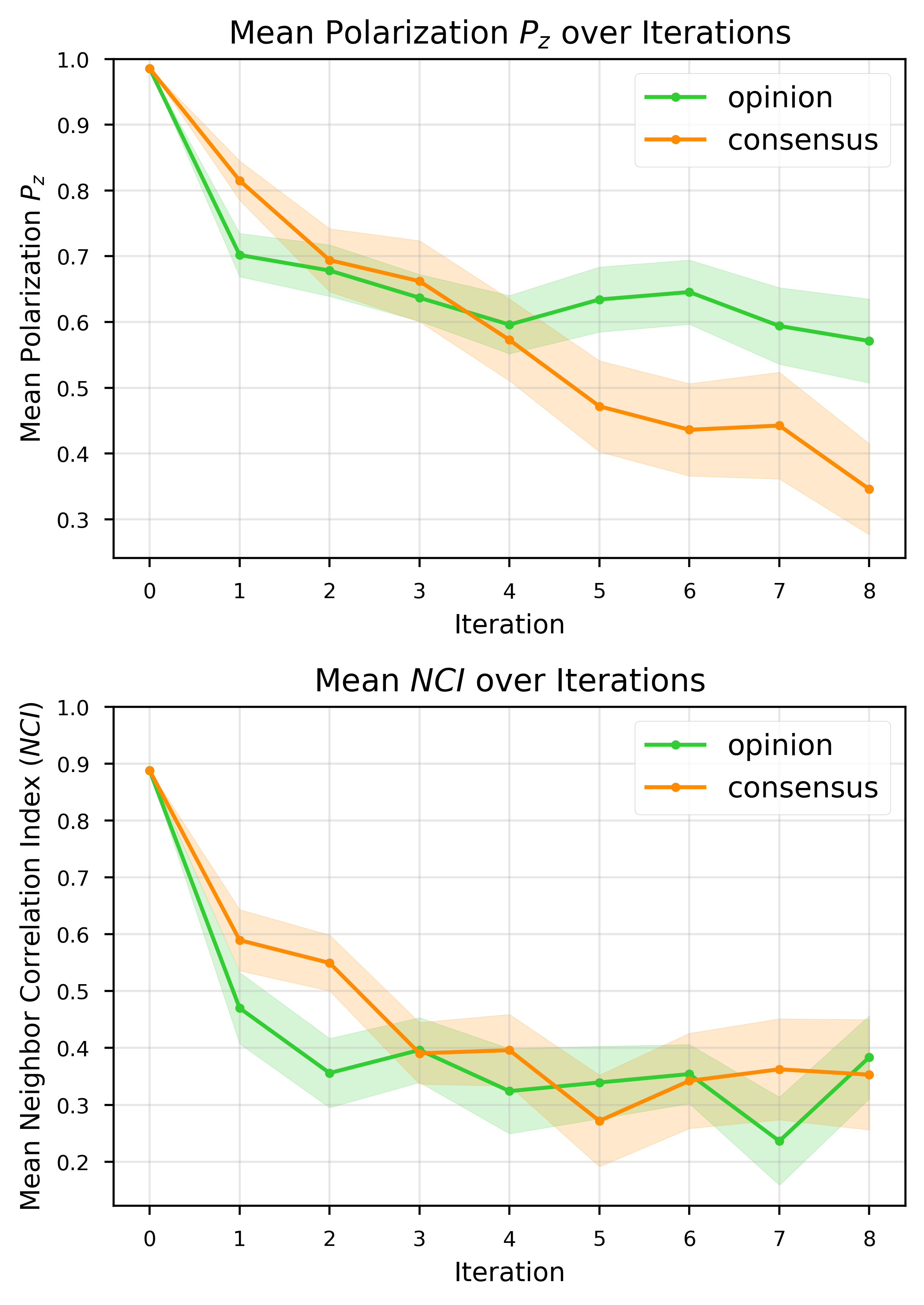}
  \end{center}
  \caption{\textbf{Experimental results}. Evolution of Mean Polarization (top) and Mean Neighbor Correlation Index (bottom) over iterations in the Grok-only condition across the two instruction framings: \textbf{opinion} (green) and \textbf{consensus} (dark orange). Shaded area  represent $+/-1$ standard error of the mean across 21 runs per condition.}
  \label{fig-5:results_llm_robustness}
\end{figure}

\subsection{Explaining Opinion Convergence in AI Networks}

To examine how task framing and prompt variations affect LLM behavior, we conducted 42 runs of the experiment using LLM-only networks. We examined two prompting conditions: (i) The \textbf{consensus} condition, described in Figure \ref{fig-1:schematics}, in which participants select and revise statements to reflect the group's likely consensus rather than their own opinion. (ii) The \textbf{opinion} condition, in which participants select and revise statements based on their own opinion.

All networks were initialized with the same question as the human and hybrid networks (see Method).

Figure \ref{fig-5:results_llm_robustness} presents the mean polarization and NCI trajectories computed on the 42 runs (21 runs per condition). Both conditions exhibited initial depolarization, but diverged substantially after iteration 4. The consensus condition achieved  lower final polarization ($M=0.36$, $CI_{95}=[0.22, 0.49]$, iteration 8) compared to the opinion condition ($M=0.57$, $CI_{95}=[0.45, 0.70]$), with a significant difference ($z=2.3$, $p=0.019$, Cohen's $d= 0.72$)   suggesting that explicitly prompting LLMs toward agreement amplifies convergence effects. 
However, the NCI in the two conditions shows a similar decreasing trend with much less differentiation, and the values at the final iteration (consensus: $M=0.33$, $CI_{95}=[0.13, 0.51]$; opinion: $M=0.38$, $CI_{95}=[0.24, 0.53]$; do not differ significantly ($z=0.49$, $p=0.62$, Cohen's $d= 0.15$). These results suggest that participants take neighboring responses into account, even when the task permits them to express their own independent opinions.

Overall, these findings might indicate that the behavior of LLM in shaping opinion dynamics is sensitive to the framing of the prompt: consensus-oriented instructions might produce stronger  depolarization outcomes compared to opinion-elicitation prompts.

\section{Discussion}

We introduce an experimental framework for studying opinion dynamics and diffusion in social networks composed of both human and AI agents. The primary contribution of this paper is the development of a fully controlled experimental setting in which collective human–AI interactions can be systematically examined. 

Our preliminary results suggest that network composition may alter the trajectory of opinion dynamics. The hybrid human-AI condition achieved the lowest final polarization while maintaining the highest neighbor agreement, consistent with recent findings that LLMs can facilitate consensus formation in deliberative settings \citep{tessler_ai_2024}. This pattern may suggest that LLM agents can function as bridges between opposing viewpoints, for example by generating statements that integrate multiple perspectives or by moderating extreme positions during the revision process \citep{ueshima2024simple}. Alternatively, improved consensus formation may arise from the injection of noise into the system \citep{shirado2017locally}, or from complementary and positive representational misalignment across different agent types \citep{huang2024characterizing,sucholutsky2023getting,shiiku_dynamics_2025}.
In contrast, the human-only condition exhibited the lowest final NCI despite moderate depolarization, which may indicate that opinion mixing occurred without the emergence of local consensus. This divergence between global convergence and local agreement merits further investigation, as it may reflect fundamentally different mechanisms of opinion updating between human and AI participants. \\
The robustness analysis of LLM-only networks revealed that prompt framing might be a factor influencing polarization outcomes. This finding carries implications for the deployment of AI agents in social platforms, such as X's Grok, where prompt design choices may inadvertently shape large-scale opinion dynamics.

\subsection{Limitations \& Future Directions}

Several limitations constrain our findings' statistical power and generalizability. First, we used only one question. Opinion dynamics likely vary across topics with different emotional valence, political salience, or scientific consensus. Future work will expand to a full set additional questions and topics. Moreover, the prompting conditions tested represent only a narrow subset of possible LLM instructions.

Second, our robustness analysis for LLMs included only 42 runs, and the experiments involving human participants were conducted only once, limiting the statistical power and scope of the analyses. Although the results indicate certain trends, future work will substantially increase the number of runs (to 100 or more) for AI-only experiments and will include additional human experiments to ensure adequate statistical power.

Third, we tested only one LLM, Grok. Given Grok’s documented history of opinion bias and recent deepfake generation following its integration with X, we considered it important to examine its behavior in opinion-diffusion dynamics and therefore deliberately chose it as the AI participant in our experiments (\cite{2026migliarini}; \cite{2026mei}; \cite{nyt-grok}). However, relying on a single model limits generalizability: our observed dynamics may differ across models with different training objectives, fine-tuning procedures, or safety constraints. Cross-model comparisons are therefore needed to establish robustness across LLMs.

Fourth, we used a limited set of relatively simple methods to characterize opinion dynamics. Matakos’ polarization index may not fully capture clustering dynamics, and our categorical stance classification ignores the semantic richness of the statements themselves. Future work could incorporate alternative polarization measures \citep{bramson2016disambiguation}, sentence embeddings to track rhetorical and argumentative change \citep{shiiku_dynamics_2025}, and continuous measures of opinion valence rather than three-category coding.

Finally, while the 5$\times$5 grid lattice provides experimental control, it oversimplifies real-world social network structures. Future work should explore more realistic topologies, such as random regular graphs or modular networks \citep{marjieh2025characterizing}, as well as weighted networks that can account for communities and influencers.

Taken together, our method has the potential to provide empirical insight into how different aspects of social cognition may be influenced by the introduction of AI agents  \citep{collins2024building}, and more broadly enabling causal understanding of the forces shaping cultural dynamics \citep{boyd1996culture}.

\section{Acknowledgments}
This work was accepted for a presentation as a talk at the Annual Conference of the Cognitive Science Society (CogSci2026) and was supported by the NSF grant ``Collaborative Research: Research Infrastructure: HNDS-I: Building Infrastructure to Study Human-AI Hybrid Societies in Experimental Social Networks'' (Award BCS-2523500). ChatGPT version 5.2 (OpenAI) was used to assist manuscript editing and proofreading. Authors reviewed each of the edit suggestions, and approved the final version.

\printbibliography

@article{williams2023epidemic,
  title={Epidemic modeling with generative agents},
  author={Williams, Ross and Hosseinichimeh, Niyousha and Majumdar, Aritra and Ghaffarzadegan, Navid},
  journal={arXiv preprint arXiv:2307.04986},
  year={2023}
}

@article{rmus2025generating,
  title={Generating computational cognitive models using large language models},
  author={Rmus, Milena and Jagadish, Akshay K and Mathony, Marvin and Ludwig, Tobias and Schulz, Eric},
  journal={arXiv preprint arXiv:2502.00879},
  year={2025}
}

@inproceedings{hashemi2025collective,
  title={Collective Social Behaviors in LLMs: An Analysis of LLMs Social Networks},
  author={Hashemi, Farnoosh and Macy, Michael},
  booktitle={Large Language Models for Scientific and Societal Advances},
  year={2025}
}

@article{breithaupt2024humans,
  title={Humans create more novelty than ChatGPT when asked to retell a story},
  author={Breithaupt, Fritz and Otenen, Ege and Wright, Devin R and Kruschke, John K and Li, Ying and Tan, Yiyan},
  journal={Scientific Reports},
  volume={14},
  number={1},
  pages={875},
  year={2024},
  publisher={Nature Publishing Group UK London}
}

@inproceedings{boyd1996culture,
  title={Why culture is common, but cultural evolution is rare},
  author={Boyd, Robert and Richerson, Peter J and others},
  booktitle={Proceedings-british academy},
  volume={88},
  pages={77--94},
  year={1996},
  organization={Oxford University Press Inc.}
}

@article{collins2024building,
  title={Building machines that learn and think with people},
  author={Collins, Katherine M and Sucholutsky, Ilia and Bhatt, Umang and Chandra, Kartik and Wong, Lionel and Lee, Mina and Zhang, Cedegao E and Zhi-Xuan, Tan and Ho, Mark and Mansinghka, Vikash and others},
  journal={Nature human behaviour},
  volume={8},
  number={10},
  pages={1851--1863},
  year={2024},
  publisher={Nature Publishing Group UK London}
}

@article{tsvetkova2024new,
  title={A new sociology of humans and machines},
  author={Tsvetkova, Milena and Yasseri, Taha and Pescetelli, Niccolo and Werner, Tobias},
  journal={Nature Human Behaviour},
  volume={8},
  number={10},
  pages={1864--1876},
  year={2024},
  publisher={Nature Publishing Group UK London}
}

@article{tessler2024ai,
  title={AI can help humans find common ground in democratic deliberation},
  author = {Michael Henry Tessler  and Michiel A. Bakker  and Daniel Jarrett  and Hannah Sheahan  and Martin J. Chadwick  and Raphael Koster  and Georgina Evans  and Lucy Campbell-Gillingham  and Tantum Collins  and David C. Parkes  and Matthew Botvinick  and Christopher Summerfield },
  journal={Science},
  volume={386},
  number={6719},
  pages={eadq2852},
  year={2024},
  publisher={American Association for the Advancement of Science}
}

@article{doshi2024generative,
  title={Generative AI enhances individual creativity but reduces the collective diversity of novel content},
  author={Doshi, Anil R and Hauser, Oliver P},
  journal={Science advances},
  volume={10},
  number={28},
  pages={eadn5290},
  year={2024},
  publisher={American Association for the Advancement of Science}
}

@article{ueshima2024simple,
  title={Simple autonomous agents can enhance creative semantic discovery by human groups},
  author={Ueshima, Atsushi and Jones, Matthew I and Christakis, Nicholas A},
  journal={Nature communications},
  volume={15},
  number={1},
  pages={5212},
  year={2024},
  publisher={Nature Publishing Group UK London}
}

@article{sucholutsky2023getting,
  title={Getting aligned on representational alignment},
  author={Sucholutsky, Ilia and Muttenthaler, Lukas and Weller, Adrian and Peng, Andi and Bobu, Andreea and Kim, Been and Love, Bradley C and Grant, Erin and Groen, Iris and Achterberg, Jascha and others},
  journal={arXiv preprint arXiv:2310.13018},
  year={2023}
}

@article{shirado2017locally,
  title={Locally noisy autonomous agents improve global human coordination in network experiments},
  author={Shirado, Hirokazu and Christakis, Nicholas A},
  journal={Nature},
  volume={545},
  number={7654},
  pages={370--374},
  year={2017},
  publisher={Nature Publishing Group UK London}
}

@article{sucholutsky2025using,
  title={Using LLMs to advance the cognitive science of collectives},
  author={Sucholutsky, Ilia and Collins, Katherine M and Jacoby, Nori and Thompson, Bill D and Hawkins, Robert D},
  journal={Nature Computational Science},
  volume={5},
  number={9},
  pages={704--707},
  year={2025},
  publisher={Nature Publishing Group US New York}
}

@article{brinkmann2023machine,
  title={Machine culture},
  author={Brinkmann, Levin and Baumann, Fabian and Bonnefon, Jean-Fran{\c{c}}ois and Derex, Maxime and M{\"u}ller, Thomas F and Nussberger, Anne-Marie and Czaplicka, Agnieszka and Acerbi, Alberto and Griffiths, Thomas L and Henrich, Joseph and others},
  journal={Nature Human Behaviour},
  volume={7},
  number={11},
  pages={1855--1868},
  year={2023},
  publisher={Nature Publishing Group UK London}
}

@article{binz2025foundation,
  title={A foundation model to predict and capture human cognition},
  author={Binz, Marcel and Akata, Elif and Bethge, Matthias and Br{\"a}ndle, Franziska and Callaway, Fred and Coda-Forno, Julian and Dayan, Peter and Demircan, Can and Eckstein, Maria K and {\'E}ltet{\H{o}}, No{\'e}mi and others},
  journal={Nature},
  pages={1--8},
  year={2025},
  publisher={Nature Publishing Group UK London}
}

@article{huang2024characterizing,
  title={Characterizing similarities and divergences in conversational tones in humans and llms by sampling with people},
  author={Huang, Dun-Ming and Van Rijn, Pol and Sucholutsky, Ilia and Marjieh, Raja and Jacoby, Nori},
  journal={arXiv preprint arXiv:2406.04278},
  year={2024}
}

@article{schmidgall2025agent,
  title={Agent laboratory: Using llm agents as research assistants},
  author={Schmidgall, Samuel and Su, Yusheng and Wang, Ze and Sun, Ximeng and Wu, Jialian and Yu, Xiaodong and Liu, Jiang and Moor, Michael and Liu, Zicheng and Barsoum, Emad},
  journal={Findings of the Association for Computational Linguistics: EMNLP 2025},
  pages={5977--6043},
  year={2025},
  publisher={Association for Computational Linguistics}
}

@article{jumper2021highly,
  title={Highly accurate protein structure prediction with AlphaFold},
  author={Jumper, John and Evans, Richard and Pritzel, Alexander and Green, Tim and Figurnov, Michael and Ronneberger, Olaf and Tunyasuvunakool, Kathryn and Bates, Russ and {\v{Z}}{\'\i}dek, Augustin and Potapenko, Anna and others},
  journal={nature},
  volume={596},
  number={7873},
  pages={583--589},
  year={2021},
  publisher={Nature Publishing Group UK London}
}

@article{marjieh2025characterizing,
  title={Characterizing the Interaction of Cultural Evolution Mechanisms in Experimental Social Networks},
  author={Marjieh, Raja and Anglada-Tort, Manuel and Griffiths, Thomas L and Jacoby, Nori},
  journal={arXiv preprint arXiv:2502.12847},
  year={2025}
}

@article{harrison2020gibbs,
  title={Gibbs sampling with people},
  author={Harrison, Peter and Marjieh, Raja and Adolfi, Federico and van Rijn, Pol and Anglada-Tort, Manuel and Tchernichovski, Ofer and Larrouy-Maestri, Pauline and Jacoby, Nori},
  journal={Advances in neural information processing systems},
  volume={33},
  pages={10659--10671},
  year={2020}
}

@misc{shiiku_dynamics_2025,
	title = {The {Dynamics} of {Collective} {Creativity} in {Human}-{AI} {Hybrid} {Societies}},
	url = {http://arxiv.org/abs/2502.17962},
	doi = {10.48550/arXiv.2502.17962},
	urldate = {2025-11-25},
	publisher = {arXiv},
	author = {Shiiku, Shota and Marjieh, Raja and Anglada-Tort, Manuel and Jacoby, Nori},
	month = may,
	year = {2025},
	note = {arXiv:2502.17962 [cs]},
}

@article{tessler_ai_2024,
	title = {{AI} can help humans find common ground in democratic deliberation},
	volume = {386},
	issn = {0036-8075, 1095-9203},
	url = {https://www.science.org/doi/10.1126/science.adq2852},
	doi = {10.1126/science.adq2852},
	language = {en},
	number = {6719},
	urldate = {2025-11-25},
	journal = {Science},
	author = {Tessler, Michael Henry and Bakker, Michiel A. and Jarrett, Daniel and Sheahan, Hannah and Chadwick, Martin J. and Koster, Raphael and Evans, Georgina and Campbell-Gillingham, Lucy and Collins, Tantum and Parkes, David C. and Botvinick, Matthew and Summerfield, Christopher},
	month = oct,
	year = {2024},
	pages = {eadq2852},
}

@misc{pournaki_conflicting_2025,
	title = {Conflicting narratives and polarization on social media},
	url = {http://arxiv.org/abs/2507.15600},
	doi = {10.48550/arXiv.2507.15600},
	urldate = {2025-11-26},
	publisher = {arXiv},
	author = {Pournaki, Armin},
	month = jul,
	year = {2025},
	note = {arXiv:2507.15600 [cs]},
}

@misc{donkers_human-agent_2025,
	title = {Human-{Agent} {Interaction} in {Synthetic} {Social} {Networks}: {A} {Framework} for {Studying} {Online} {Polarization}},
	shorttitle = {Human-{Agent} {Interaction} in {Synthetic} {Social} {Networks}},
	url = {http://arxiv.org/abs/2502.01340},
	doi = {10.48550/arXiv.2502.01340},
	urldate = {2025-12-03},
	publisher = {arXiv},
	author = {Donkers, Tim and Ziegler, Jürgen},
	month = jun,
	year = {2025},
	note = {arXiv:2502.01340 [physics]},
}

@misc{wang_decoding_2024,
	title = {Decoding {Echo} {Chambers}: {LLM}-{Powered} {Simulations} {Revealing} {Polarization} in {Social} {Networks}},
	copyright = {Creative Commons Attribution 4.0 International},
	shorttitle = {Decoding {Echo} {Chambers}},
	url = {https://arxiv.org/abs/2409.19338},
	doi = {10.48550/ARXIV.2409.19338},
	urldate = {2025-12-03},
	publisher = {arXiv},
	author = {Wang, Chenxi and Liu, Zongfang and Yang, Dequan and Chen, Xiuying},
	year = {2024},
	note = {Version Number: 2},
}

@article{zhao_evolution_2023,
	title = {The {Evolution} of {Polarization} in {Online} {Conversation}: {Twitter} {Users}’ {Opinions} about the {COVID}-19 {Pandemic} {Become} {More} {Politicized} over {Time}},
	volume = {2023},
	copyright = {https://creativecommons.org/licenses/by/4.0/},
	issn = {2578-1863},
	shorttitle = {The {Evolution} of {Polarization} in {Online} {Conversation}},
	url = {https://www.hindawi.com/journals/hbet/2023/9094933/},
	doi = {10.1155/2023/9094933},
	language = {en},
	urldate = {2025-12-15},
	journal = {Human Behavior and Emerging Technologies},
	author = {Zhao, Weize and Walasek, Lukasz and Brown, Gordon D. A.},
	editor = {Yan, Zheng},
	month = jul,
	year = {2023},
	pages = {1--14},
}

@article{macy_opinion_2019,
	title = {Opinion cascades and the unpredictability of partisan polarization},
	volume = {5},
	issn = {2375-2548},
	url = {https://www.science.org/doi/10.1126/sciadv.aax0754},
	doi = {10.1126/sciadv.aax0754},
	language = {en},
	number = {8},
	urldate = {2026-01-05},
	journal = {Sci. Adv.},
	author = {Macy, Michael and Deri, Sebastian and Ruch, Alexander and Tong, Natalie},
	month = aug,
	year = {2019},
	pages = {eaax0754},
}

@article{priniski_network_2026,
	title = {Network structure shapes consensus dynamics through individual decisions},
	volume = {123},
	issn = {0027-8424, 1091-6490},
	url = {https://pnas.org/doi/10.1073/pnas.2520483123},
	doi = {10.1073/pnas.2520483123},
	language = {en},
	number = {2},
	urldate = {2026-01-09},
	journal = {Proc. Natl. Acad. Sci. U.S.A.},
	author = {Priniski, J. Hunter and Linford, Bryce and Hirschmann, Anna and Venumuddala, Sai Krishna and Morstatter, Fred and Rodriguez, Nancy and Brantingham, P. Jeffrey and Lu, Hongjing},
	month = jan,
	year = {2026},
	pages = {e2520483123},
}

@article{thurner_why_2025,
	title = {Why more social interactions lead to more polarization in societies},
	volume = {122},
	issn = {0027-8424, 1091-6490},
	url = {https://pnas.org/doi/10.1073/pnas.2517530122},
	doi = {10.1073/pnas.2517530122},
	language = {en},
	number = {44},
	urldate = {2026-01-12},
	journal = {Proc. Natl. Acad. Sci. U.S.A.},
	author = {Thurner, Stefan and Hofer, Markus and Korbel, Jan},
	month = nov,
	year = {2025},
	pages = {e2517530122},
}

@inproceedings{chitra_analyzing_2020,
	address = {Houston TX USA},
	title = {Analyzing the {Impact} of {Filter} {Bubbles} on {Social} {Network} {Polarization}},
	isbn = {978-1-4503-6822-3},
	url = {https://dl.acm.org/doi/10.1145/3336191.3371825},
	doi = {10.1145/3336191.3371825},
	language = {en},
	urldate = {2026-01-13},
	booktitle = {Proceedings of the 13th {International} {Conference} on {Web} {Search} and {Data} {Mining}},
	publisher = {ACM},
	author = {Chitra, Uthsav and Musco, Christopher},
	month = jan,
	year = {2020},
	pages = {115--123},
}

@article{garimella_quantifying_2018,
	title = {Quantifying {Controversy} on {Social} {Media}},
	volume = {1},
	copyright = {https://www.acm.org/publications/policies/copyright\_policy\#Background},
	issn = {2469-7818, 2469-7826},
	url = {https://dl.acm.org/doi/10.1145/3140565},
	doi = {10.1145/3140565},
	language = {en},
	number = {1},
	urldate = {2026-01-13},
	journal = {Trans. Soc. Comput.},
	author = {Garimella, Kiran and Morales, Gianmarco De Francisci and Gionis, Aristides and Mathioudakis, Michael},
	month = mar,
	year = {2018},
	pages = {1--27},
}

@article{cinus_effect_2022,
	title = {The {Effect} of {People} {Recommenders} on {Echo} {Chambers} and {Polarization}},
	volume = {16},
	issn = {2334-0770, 2162-3449},
	url = {https://ojs.aaai.org/index.php/ICWSM/article/view/19275},
	doi = {10.1609/icwsm.v16i1.19275},
	urldate = {2026-01-16},
	journal = {ICWSM},
	author = {Cinus, Federico and Minici, Marco and Monti, Corrado and Bonchi, Francesco},
	month = may,
	year = {2022},
	pages = {90--101},
}

@article{matakos_measuring_2017,
	title = {Measuring and moderating opinion polarization in social networks},
	volume = {31},
	issn = {1384-5810, 1573-756X},
	url = {http://link.springer.com/10.1007/s10618-017-0527-9},
	doi = {10.1007/s10618-017-0527-9},
	language = {en},
	number = {5},
	urldate = {2026-01-19},
	journal = {Data Min Knowl Disc},
	author = {Matakos, Antonis and Terzi, Evimaria and Tsaparas, Panayiotis},
	month = sep,
	year = {2017},
	pages = {1480--1505},
}

@article{friedkin1990social,
  title={Social influence and opinions},
  author={Friedkin, Noah E and Johnsen, Eugene C},
  journal={Journal of mathematical sociology},
  volume={15},
  number={3-4},
  pages={193--206},
  year={1990},
  publisher={Taylor \& Francis}
}

@article{deffuant2000mixing,
  title={Mixing beliefs among interacting agents},
  author={Deffuant, Guillaume and Neau, David and Amblard, Frederic and Weisbuch, G{\'e}rard},
  journal={Advances in Complex Systems},
  volume={3},
  number={01n04},
  pages={87--98},
  year={2000},
  publisher={World Scientific}
}

@article{acemoglu2011opinion,
  title={Opinion dynamics and learning in social networks},
  author={Acemoglu, Daron and Ozdaglar, Asuman},
  journal={Dynamic Games and Applications},
  volume={1},
  number={1},
  pages={3--49},
  year={2011},
  publisher={Springer}
}

@article{bramson2016disambiguation,
  title={Disambiguation of social polarization concepts and measures},
  author={Bramson, Aaron and Grim, Patrick and Singer, Daniel J and Fisher, Steven and Berger, William and Sack, Graham and Flocken, Carissa},
  journal={The Journal of Mathematical Sociology},
  volume={40},
  number={2},
  pages={80--111},
  year={2016},
  publisher={Taylor \& Francis}
}

@article{2026migliarini,
    author = {Migliarini, Matteo and Ercevik, Berat and Olowe, Oluwagbemike and Fatima, Saira and Zhao, Sarah and Le, Minh Anh and Sharma, Vasu and Panda, Ashwinee},
    title = {@GrokSet: multi-party Human-LLM Interactions in Social Media},
    journal = {arXiv preprint arxiv:2602.21236},
    year = {2026}
}

@article{2026mei,
    author = {Mei, Katelyn Xiaoying and Wolfe, Robert and Weber, Nicholas and Saveski, Martin},
    title = {Grok in the Wild: Characterizing the Roles and Uses of Large Language Models on Social Media},
    journal = {arXiv preprint arXiv:2602.11286},
    year = {2026} 
}

@article{nyt-grok,
    author = {Conger, Kate and Freedman, Dylan and Thompson, Stuart A.} ,
    title = {Musk’s Chatbot Flooded X With Millions of Sexualized Images in Days, New Estimates Show},
    journal = {The New York Times},
    year = {2026},
    month = {January},
    note = {Accessed: 2026-05-05}
}

\end{document}